\documentclass[a4paper,11pt]{article}
\usepackage{aaskaiid}
\usepackage{aas_macros}
\usepackage{soul}

\newcommand{\hi}{\mbox{H\,{\sc i}}}

\newcommand{\degree}{\ensuremath{^\circ}}

\definecolor{green}{rgb}{0,0.4,0}

\title{\hi\ Galaxy Science with the SKA}
\ShortTitle{HI surveys with the SKA}

\author[1]{Jing Wang}
\affiliation[1]{Kavli Institute for Astronomy and Astrophysics, Peking University, Beijing 100871, People's Republic of China }
\emailAdd{jwang\_astro@pku.edu.cn}

\author[2]{D.J. Pisano}
\affiliation[2]{University of Cape Town, Private Bag X3, Rondebosch, 7701, Republic of South Africa}
\emailAdd{dj.pisano@uct.ac.za}

\author[2]{Sarah Blyth}
\emailAdd{sarah.blyth@uct.ac.za}

\author[3]{Neeraj Gupta}
\affiliation[3]{Inter-University Centre for Astronomy and Astrophysics, Post Bag 4, Ganeshkhind, Pune 411 007, India}
\emailAdd{ngupta@iucaa.in}

\author[4]{Barbara Catinella}
\affiliation[4]{International Centre for Radio Astronomy Research (ICRAR), The University of Western Australia, 35 Stirling Highway, Crawley, WA 6009, Australia }
\emailAdd{barbara.catinella@uwa.edu.au}

\author[4]{Lister Staveley-Smith}
\emailAdd{Lister.Staveley-Smith@uwa.edu.au}

\author[5]{Paolo Serra}
\affiliation[5]{INAF – Osservatorio Astronomico di Cagliari, Via della Scienza 5, I-09047 Selargius, (CA), Italy }
\emailAdd{paolo.serra@inaf.it}

\author[6,7]{Elizabeth A. K. Adams}
\affiliation[6]{Netherlands Institute for Radio Astronomy (ASTRON), Oude Hoogeveensedijk 4, 7991 PD Dwingeloo, the Netherlands}
\affiliation[7]{Kapteyn Astronomical Institute, University of Groningen Postbus 800, 9700 AV Groningen, the Netherlands}
\affiliation[8]{SKA Observatory, SKA-Mid Science Operation Centre, 2 Fir St, Observatory, Cape Town 7925, South Africa}
\emailAdd{adams@astron.nl}

\author[6,2,7]{W.J.G. de Blok}
\emailAdd{blok@astron.nl}

\author[4]{Martin Meyer}
\emailAdd{martin.meyer@uwa.edu.au}

\author[9]{Lourdes Verdes-Montenegro}
\affiliation[8]{Instituto de Astrofísica de Andalucía, Glorieta de Astronomía, Granada IAA-CSIC, Spain }
\emailAdd{lourdes@iaa.es}

\author[6,7]{Tom Oosterloo}
\emailAdd{oosterloo@astron.nl}

\abstract{This chapter introduces the contributions of the \hi\ galaxy science in this volume reviewing the latest developments and urgent questions in \hi\ galaxy science, providing guiding principles for a layered set of future key science projects. The key science will include: a complete censuses of \hi\ morphologies and kinematics at sub-kpc and 1 km/s resolution within and around galaxies in the nearby Universe; a measurement of the cosmic \hi\ mass density and \hi\ mass function evolution at least up to $z\sim1$; an improved understanding of the Universe at $z>1$, particularly the balance between cold molecular and cool atomic gas. We also provide a view of the synergistic multi-wavelength surveys  available in 2028+ in the southern hemisphere. This effort will improve our understanding of the baryon cycle across a significant fraction of the cosmic history, including the processes of gas accretion, consumption and removal as well as AGN and star formation feedback. Based on these science goals, the earlier proposed three-tiered survey strategy remains, but survey parameters and predictions are adjusted according to AA* and AA4 developments. This  chapter is an update of the earlier "Advancing Astrophysics with the Square Kilometre Array" chapter `\hi\ Science with the SKA' by Staveley-Smith \& Oosterloo.}

\begin{document}
\maketitle

\section{Introduction}
Neutral atomic hydrogen (\hi) serves as the fundamental building block of galaxies and is the primary reservoir for star formation throughout cosmic time. As the most abundant element in the Universe and the raw material from which molecular clouds and subsequently stars form, understanding the distribution, kinematics, and physical conditions of atomic hydrogen is essential for a complete picture of galaxy formation and evolution. The study of \hi\ in galaxies has undergone a revolutionary transformation over the past few decades, evolving from single-dish surveys that mapped the coarse global properties of nearby galaxies to sophisticated interferometric observations that resolve the intricate spatial and kinematic structures of gas within and around galaxies. These advances have revealed that \hi\ is not merely a passive tracer of galactic potential wells, but rather a dynamic component intimately connected to the baryonic cycle of gas accretion and removal, star formation, and feedback that drives galaxy evolution.

The emerging era of Square Kilometre Array (SKA) facilities promises to push these boundaries even further, offering unprecedented sensitivity, angular resolution, and survey speed that will transform our understanding of \hi\ in cosmic time. The SKA will enable the detection of \hi\ in galaxies out to redshifts where the cosmic star formation rate peaked, map the large-scale distribution of neutral gas in the cosmic web, and resolve the internal gas dynamics of galaxies with exquisite detail. These capabilities arrive at a critical juncture in astrophysics, when theoretical models of galaxy formation within the hierarchical framework of cold dark matter cosmology are mature enough to make detailed predictions about the \hi\ content and distribution of galaxies, yet face persistent challenges in accurately reproducing observed \hi\ properties and their scaling relations.

Just over ten years ago, in  \citet[][SO15 hereafter]{SO15}, the first straw-man surveys were designed to advance \hi\ galaxy studies to the next level. 
The 1000-hr fiducial (and 10,000-hr commensal) \hi\ surveys introduced at that time were largely based on the re-baselined SKA-mid telescope, which consisted of 133 15-m diameter SKA dishes and 64 13.5-m diameter MeerKAT dishes. This version is still the goal, but is referred to as Array Assembly 4 (AA4) in this volume. For fiducial \hi\ emission surveys, AA4 baselines up to about 30~km are useful, resulting in an angular resolution of  $2^{\prime\prime}$ for deep surveys at $z \approx 0.5$. The shallower surveys with a wide area result in resolutions of $5^{\prime\prime}$ to $10^{\prime\prime}$. However, the more immediate goal is the so-called AA* configuration, consisting of 80 SKA dishes and 64 MeerKAT dishes. The SKA dishes include 16 from MeerKAT extension (MeerKAT +) of a similar diameter and design.

This volume brings together contributions from leading experts in \hi\ astronomy, addressing key topics that span from the detailed physics of multi-phase gas in the Milky Way and nearby galaxies to the cosmological distribution of \hi\ in the cosmic web. The chapters collectively explore how deep observations can reveal the complex processes of gas inflow and outflow that regulate galaxy growth, how resolved \hi\ kinematics can constrain fundamental physics including the nature of dark matter and gravity, how absorption line studies provide a redshift-independent probe of cold gas from the local Universe to the early Universe, and how next-generation radio observations will be synergistically combined with optical spectroscopy and cosmological simulations. Together, these contributions provide a comprehensive snapshot of the field on the eve of the SKA era, highlighting both the tremendous scientific potential of forthcoming facilities and the theoretical and observational foundations upon which future progress will be built.

\section{\hi\ science case}
The following summaries provide an overview of each chapter's contributions to this broader narrative of understanding \hi\ galaxies in the context of galaxy formation and evolution. Although many cover the same broad topics as in 2015, results from the pathfinder telescopes have advanced the field significantly in the past decade.   

These chapters are organized to trace \hi\ through the galactic baryonic cycle, progressively moving from small to large scales, and linking to galaxy evolution. Accretion and feedback mechanisms drive gas flows \citep{deBlok01.2026.SKA} and shape the structure of the interstellar medium \citep{Miville-Deschenes01.2026.SKA}, establishing the conditions and providing the fuel from which stars form \citep{Rosolowsky01.2026.SKA}. These processes unfold within the immediate-and arguably most critical—environment of the host dark matter halo \citep{Lelli01.2026.SKA}, with a specific focus on the ongoing puzzles at the extremely faint end of the galaxy distribution \citep{Deg01.2026.SKA}. Traced by resolved kinematic features, the perspective shifts outward to the larger environments of galaxy groups \citep{Ramatsoku01.2026.SKA} and the cosmic web \citep{HengxingPan01.2026.SKA}. Finally, the narrative looks toward the more distant Universe, exploring cosmic evolution through absorption-line probing \citep{Mahony01.2026.SKA}, synergy with optical spectroscopy \citep{Duncan01.2026.SKA}, and comparisons with theoretical modeling \citep{Lagos01.2026.SKA}.   


\subsection{Deep observations of cold gas inflow and outflow \citep{deBlok01.2026.SKA}}
\citet{deBlok01.2026.SKA} focus on the baryon cycle, specifically the detection and characterization of gas accretion and feedback mechanisms. They emphasize that detecting the faint, low-column density \hi\ in the circumgalactic medium is crucial to understand how galaxies sustain star formation over cosmic time. By leveraging the high sensitivity of the SKA, researchers will be able to resolve gas inflows from the intergalactic medium and quantify the energetics of outflows driven by stellar and AGN feedback, providing a direct observational link to the processes that regulate galaxy growth.

\subsection{The multi-phase \hi\ of the Milky Way and nearby galaxies \citep{Miville-Deschenes01.2026.SKA}}
Focusing on the internal physics of the interstellar medium, \citet{Miville-Deschenes01.2026.SKA} examined the transition between the cold neutral medium and the warm neutral medium. They discuss new phase-separation techniques that use spectral information to resolve the thermodynamic structure of \hi\ in the Milky Way and Magellanic Clouds. These high-resolution studies are essential for understanding how the local environment and turbulence shape the gas reservoirs that eventually collapse to form molecular clouds and stars.

\subsection{From Atomic Gas to Star Formation \citep{Rosolowsky01.2026.SKA}}
\citet{Rosolowsky01.2026.SKA} discuss the lifecycle of gas by examining the transition from atomic \hi\ to molecular $H_2$ and subsequent star formation. They review the Kennicutt-Schmidt law and how the efficiency of star formation is linked to the properties of the atomic gas reservoir. With the SKA's ability to resolve these structures at parsec scales in nearby galaxies, researchers will finally be able to disentangle the complex interplay between gas cooling, cloud formation, and the stellar feedback that regulates the birth of new stars

\subsection{Testing dark matter models and modified gravity theories with spatially resolved \hi\ observations \citep{Lelli01.2026.SKA}}
\citet{Lelli01.2026.SKA} discuss \hi\ as a tool for fundamental physics, specifically testing the nature of dark matter and modified gravity theories. Spatially resolved \hi\ rotation curves provide some of the strongest constraints on the distribution of mass in galaxies. The chapter addresses the ``rotation-curve diversity'' problem—where galaxies of similar mass exhibit varying inner density profiles—and evaluates how these observations can distinguish between cold dark matter models with baryonic feedback and alternative theories like Modified Newtonian Dynamics (MOND).

\subsection{The Bright Future of the Dark and Dim Universe \citep{Deg01.2026.SKA}}
\citet{Deg01.2026.SKA} explore the low-mass frontier of galaxy formation, focusing on two complementary populations: ultra-diffuse galaxies (UDGs) and ``dark'' galaxies that may lack significant stellar components. The chapter highlights how the SKA-Mid sensitivity will revolutionize our understanding of the \hi\ mass function at the lowest masses, where current surveys are limited. These observations are vital for addressing the ``missing satellites'' and ``too big to fail'' problems by determining whether the scarcity of small galaxies is due to a lack of gas or a failure of star formation in low-mass dark matter halos.

\subsection{Chapter 6: Resolved \hi\ and Environmental Dynamics \citep{Ramatsoku01.2026.SKA}}
\citet{Ramatsoku01.2026.SKA} highlight the impact of local environments, such as galaxy groups and clusters, on resolved \hi\ properties. They detail the dynamical processes, including ram-pressure stripping and tidal interactions, that remove or redistribute gas within galaxies. By resolving these features, the SKA will provide a detailed look at how environmental quenching occurs and how the gaseous disks of galaxies are transformed as they fall into denser regions of the Universe.

\subsection{Studying \hi\ and the Cosmic Web in the era of SKA \citep{HengxingPan01.2026.SKA}}
There are clear connections between galactic gas content and the large-scale structure of the Universe. \citet{HengxingPan01.2026.SKA} explore how the SKA will map \hi\ within the cosmic web, revealing the pathways through which gas is accreted onto galactic disks. By surveying large volumes with high column density sensitivity, the SKA will allow researchers to understand how the position of a galaxy within the cosmic web—whether in a filament, node, or void—governs its evolution and gas processing.

\subsection{\hi\ 21-cm absorption studies with the SKA \citep{Mahony01.2026.SKA}}
Observations of the \hi\ 21-cm line in absorption provide a redshift-independent and dust-unbiased probe of cold gas across cosmic time. Unlike emission surveys, the sensitivity of absorption measurements does not degrade with distance, making it uniquely powerful for tracing cold neutral gas from the local Universe to ``cosmic noon'' ($z \sim 1-3$) with SKA-mid and beyond with SKA-low. The chapter of \citet{Mahony01.2026.SKA} highlights the potential of the SKA to identify neutral gas outflows from luminous AGN, trace the evolution of the cold gas fraction in and around normal galaxies, and study the ``21-cm forest'' toward high-redshift radio sources as a probe of neutral IGM in the early Universe, providing critical constraints on the evolution of neutral gas and the impact of AGN feedback across cosmic time.

\subsection{Unlocking the full potential of SKAO extra-galactic science with high-multiplex optical spectroscopy \citep{Duncan01.2026.SKA}}
Recognizing that radio observations do not exist in a vacuum, \citet{Duncan01.2026.SKA} discuss the essential synergies between the SKAO and next-generation multi-object spectrographs. Optical spectroscopy provides the precise redshifts and host galaxy properties (such as stellar mass and metallicity) necessary to contextualize \hi\ detections. This multi-wavelength approach allows for the separation of star-formation and AGN activity, enabling a holistic view of how the gaseous reservoirs traced by the SKA relate to the chemical and kinematic histories of galaxies.

\subsection{Cosmological Galaxy Formation Modeling in the Era of the Square Kilometre Array \citep{Lagos01.2026.SKA}}
\citet{Lagos01.2026.SKA} review the state-of-the-art in cosmological simulations, bridging the gap between sub-parsec star formation and gigaparsec cosmic structures. They argue for a ``wedding-cake'' strategy that combines high-resolution hydrodynamical simulations with large-volume semi-analytic models to interpret SKA data. A key emphasis is placed on ``forward modeling'', where simulations are used to create mock observables that can be directly compared with radio interferometric data, ensuring that theoretical models can both guide and be constrained by future discoveries.

\section{Pioneering results on evolution of HI with redshift and future prospects}
One important science goal of SKA that is not presented in the above chapters is tracing \hi\ back to the middle age of Universe, investigating the evolution of the \hi\ content of galaxies in different environments over cosmic time. 
Since the 2015 book, some pioneering results have appeared.
Over the past ten years, steady progress has been made in detecting \hi\ in galaxies beyond the local Universe thanks to various surveys coming online on SKA pathfinder instruments such as ASKAP, uGMRT, MeerKAT, FAST, and VLA with wide-band low-frequency capabilities. 
However, substantial work is still needed before we are able to converge to a consensus picture. Below, we briefly summarize the status.

Deep surveys such as CHILES \citep[e.g.][]{Luber_2025}, GMRT-CATz1 \citep{Chowdhury_2022}, MIGHTEE-HI \citep[e.g.][]{Jarvis_2025} and LADUMA \citep[e.g.][]{Blyth_16} are building on the work of the single dish HIGHz \citep{Catinella_2015} and interferometric BUDHIES \citep{Verheijen_2007} surveys to increase the sample of direct \hi\ detections in galaxies at $z>0.2$. 
Just in the past two years, these samples have increased dramatically. \cite{Jarvis_2025} recently discovered 11 galaxies with $z=0.26-0.3841$ in a subset of the MIGHTEE-\hi\ survey data.  \cite{BlueBird_26} reported four new detections in CHILES with $z=0.22-0.47$. Finally, \cite{Xi_2024} report 29 detections between $z=0.2-0.4$ using FAST.  While these samples are still small (order tens), current deep surveys are demonstrating the feasibility of detecting, and even resolving, individual galaxies at intermediate redshifts. Further exciting results of direct detections at higher redshift are coming from gravitationally lensed systems, with the record-holding detection so far in a star-forming galaxy in the Dragon Arc at $z=0.725$ by \cite{Lawrie_2025}. 

Although pathfinder surveys have also been increasing the sample sizes of spatially resolved direct \hi\ detections, these samples are still poorly resolved. This means that detailed studies of resolved \hi\ kinematics of galaxies are still limited to small samples of higher luminosity, more extended galaxies. Due to the still-limited numbers of direct detections for $z>0.2$, key science aims such as probing the \hi\ mass function (HIMF) and the cosmic neutral hydrogen density, $\Omega_{\rm{HI}}$, as well as the evolution of the \hi\ scaling relations at intermediate to high redshifts, are currently being addressed using statistical methods based on \hi\ stacking. 

Investigating how the gas content of galaxies changes compared to their stellar mass over cosmic time can provide valuable information on the build-up of mass in galaxies and therefore on their evolution. The so-called \hi\ -scaling relations typically compare galaxies' \hi\ masses to their stellar masses or a suitable proxy, for example, B-band magnitude, and are also useful for broad comparison to predictions from cosmological simulations. Enabled by the use of the \hi\ stacking technique, these average relations have recently been measured for the first time using surveys on pathfinder telescopes for samples of star forming galaxies beyond the local Universe \citep{Guo_2021,Sinigaglia_2022,Rhee_2022, Chowdhury_2022, Bianchetti_2025, Luber_2025, Depalma2025}. There is some tension in these results, however, with different studies measuring different evolution in the normalisation and slope of the $M_{HI}-M_{*}$ scaling relation from z=0 \citep{Fabello11,Guo_2021,Rhee_2022}, 
through $z\sim0.35$ \citep{Bianchetti_2025, Bera_2023} out to $z\sim1$ \citep{Chowdhury_2022}.  These differences could be the result of different sample selection between studies, cosmic variance, or differences in the stacking techniques.  
Similar discrepancies arise when using \hi\ stacking and scaling relations to infer the \hi\ mass function across the same redshift range \citep{Bera_2022, Chowdhury_2024, Pan_2025, Sinigaglia_2025}.   

The combination of deep \hi\ surveys and the \hi\ stacking technique has also enabled measurements of the cosmic neutral gas density, $\Omega_\mathrm{HI}$, across intermediate redshift \citep{Rhee_2016,Bera_2019,Bera_2023,Sinigaglia_2025} out to $z\sim1$ \citep{Chowdhury_2020} where previous results relied on damped Lyman-alpha absorber measurements. It is still difficult to discern clear trends over the intermediate redshift range as there is still some scatter between the various results, again due to sample selection and cosmic variance. However the results seem to be generally consistent, within uncertainties, with the absorber measurements. Difficulties in measuring $\Omega_\mathrm{HI}$ using stacking at these intermediate redshifts include uncertainties on the \hi\ scaling relations (determined from stacking as well) employed in the estimations as well as the need for large and complete optical redshift catalogues. 

\subsection{Outlook}
A deep \hi\ survey with AA* in Band 1 will significantly increase the number of direct \hi\ detections at $z>0.2$. This will enable direct measurements of the HIMF using individual source counts at intermediate to high redshifts which can be compared with the current best estimates based on the various statistical methods. Measurements of the HIMF with direct source counts from $0<z<1.5$ will also enable more direct and systematic determination of the evolution of the cosmic neutral hydrogen density, $\Omega_{HI}$, to be compared to the current \hi\ stacking results. 

Moreover, a large sample of direct detections at $z>0.2$ using SKA will also enable a comparison of the directly measured mean and median \hi\ scaling relations to the current state-of-the-art stacking results in the same redshift range, helping to determine the sources of differences in the current results. However, it will be important to aim for large enough cosmic volumes such that the effects of cosmic variance do not dominate the results.

\section{Possible \hi\ surveys}

SO15 adopted the three-tiered "wedding cake" approach commonly used in telescope surveys for galaxies, which satisfy different scientific needs in a balanced way.
Here we follow the same approach, but provide updated survey predictions that incorporate the AA* and AA4 configurations alongside the latest theoretical models of HI evolution. Crucially, our estimates now account for RFI (Radio Frequency Interference) contamination based on recent monitoring and observational data from MeerKAT.

The survey parameters and predictions are summarized in Table 1.
The survey designs (columns 1-3) remain consistent with SO15, comprising three 1,000-hr surveys (Tab 1 in SO15), and three 10,000-hr surveys (Tab 2 in SO15). Most cosmological redshift dependent calculations are based on equations from \citet{Meyer_17}.

For prediction with AA* and AA4, the sensitivity and resolution estimates assume the all dish array pointing toward ($\alpha$, $\delta$)= (60$\degree$, -30$\degree$).
The (robust, taper) sets of (0,1), (natural, 1), (1, 4), and (0, 16) are employed, to approximate the resolutions of 2, 5, 10, and 15'' (see de Blok et al., Chapter 1).
The predicted redshift range and detectable galaxies are estimated with the rms based \hi\ mass limit, $M_\text{HI,lim}$ (Figure 1-a), assuming galaxy sizes following the size-mass relation of \citet{Wang16}, $5\sigma$ detection threshold per beam and 100 km/s line width. They also assume redshift dependent HIMF predicted by GAEA (GAlaxy Evolution and Assembly) Semi Analytical Modeling (SAM, SMI \& SMII, \citealt{DeLucia_24}), and realistic RFI flagging rate for each redshift interval (Figure 1-b).

Compared to the values in SO15, the most important change is the significant increase in detectable galaxy numbers in all surveys, possibly due to the improved antenna sensitivity and better characterized RFI environment.
We note that better $N_\mathrm{HI,lim}$ sensitivities can be achieved with reasonable smoothing (see Chapter 1 for more details), and higher redshifts (maximal $z=3$ as allowed by band-1) can be reached through stacking or direct detection of rare galaxies in each survey.

\begin{figure}[h]
    \centering
       \includegraphics[width=0.49\textwidth]{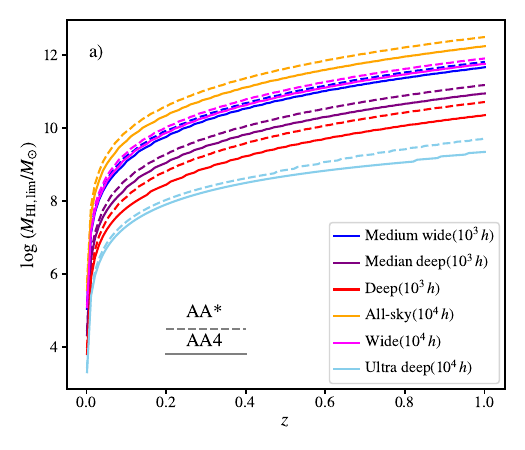}
       \includegraphics[width=0.49\textwidth]{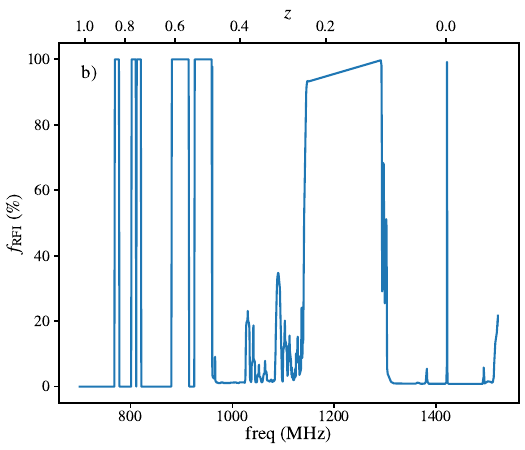}
    \caption[]{The \hi\ mass limit ($M_\mathrm{HI,lim}$) as a function of redshift and RFI contaminating rate ($f_\mathrm{RFI}$) as a function of frequency. As labeled in panel a, the different colors represent the two sets (1,000 and 10,000 $h$) of three-tiered surveys with increasing depth and decreasing sky areas (Table 1); the solid and dashed line styles are for AA4 and AA* surveys respectively.  In panel b, for the frequency range $>$960 MHz, we assume $f_\mathrm{RFI}$ from fig A1 of \citet{Heywood2024}, while for the range below 960 MHz we assume $f_\mathrm{RFI}=1$ for the RFI windows suggested in SARAO knowledgebase.\footnote{\url{https://skaafrica.atlassian.net/wiki/spaces/ESDKB/pages/305332225/Radio+Frequency+Interference+RFI}}
     It is clear that some redshift ranges are inaccessible, which most strongly affect the SKA medium-deep surveys.
    }
    \label{fig:rfi}
\end{figure}

\begin{table}[htbp]

\centering
\caption{Survey parameters}
\label{tab:survey}
\begin{tabular}{@{} l c c c c c c c @{}}

\hline
Survey & $\Omega$  & Frequency & beam & $N$ & $\langle z \rangle$ ($z_{\text{lim}}$) & $N_{\text{HI,lim}}$ & Chapter \\
       &(deg$^2$)  & (MHz)     &            &       &         &($10^{20}$ cm$^{-2}$) & \\
(1) & (2) & (3) & (4)  &(5) & (6) & (7) &(8) \\

\hline


\hline
\multicolumn{8}{c}{AA*} \\
\hline
\multicolumn{8}{c}{1,000-h} \\
Medium wide  & 400      & 950--1420 & 11" & 57,300  &0.06 (0.21) & 1.5   & 2, 3, 4 \\
Medium deep  & 20       & 950--1420 & 4.5" &  16,000  &0.19 (0.38)  & 2.4 & 1, 6,7  \\
Deep         & 1        & 600--1050 & 2.4" &  3,500   &0.29  (0.57)    & 3.4 &   9 \\
\multicolumn{8}{c}{10,000-h} \\
All-sky      & 20,000   & 950--1420 & 17" &  617,400 & 0.05 (0.11) & 2.1  &  8, 9\\
Wide         & 5,000    & 950--1420 & 11" &  594,600 & 0.06 (0.20)  & 1.7  & 1, 5,10 \\
Ultra deep   & 1        & 450--1050 & 2.4"  & 36,500  & 0.50 (1.00)    & 1.6 & 1,9 \\

\hline
\hline
\multicolumn{8}{c}{AA4} \\
\hline

\multicolumn{8}{c}{1,000-h} \\
Medium wide  & 400      & 950--1420 & 10.4" & 74,000  &0.07 (0.25) & 1.4   &  2, 3, 4 \\
Medium deep  & 20       & 950--1420 & 4.5" &  34,600  &0.25 (0.47)  & 2.1 & 1, 6, 7 \\
Deep         & 1        & 600--1050 & 2.7" &  9,000   &0.36  (0.76)    & 2.2 &  9 \\
\multicolumn{8}{c}{10,000-h} \\
All-sky      & 20,000   & 950--1420 & 17.6" & 1,174,300 & 0.05 (0.14) & 1.5   &  8, 9\\
Wide         & 5,000    & 950--1420 & 10.4" & 788,900 & 0.06 (0.23) & 1.5 & 1, 5, 10  \\
Ultra deep   & 1        & 450--1050 & 2.7"  & 56,300  & 0.58 (1.00)    & 1.0 & 1, 9 \\

\hline
\end{tabular}
 {\textbf{Note:} The survey configurations (coloumns 1-3) are consistent with those in Tables 1 and 2 of SO15. For each stage (AA* and AA4), there are three 1,000 h surveys (first three rows), and three 10,000 h surveys (the remaining three rows).
(4) The beam size (mean of bmaj and bmin).
(5) The number of detectable galaxies.
(6) The average redshift, $\langle z \rangle$, is calculated as the detectable galaxy number weighted average redshift in the range 0 to $z_\text{lim}$. The redshift limit, $z_\text{lim}$, is estimated as the highest $z$ where $M_\text{HI,lim}=10^{10}\,M_{\odot}$. 
$M_\text{HI,lim}$ is estimated as $5\sigma$ limit, assuming 100 km$/$s line width.}
\end{table}

\subsection{Design considerations}

\subsubsection{Multiwavelength data}
The landscape of extragalactic astronomy has changed dramatically in the last 15 years. By the time SKA-mid begins delivering \hi\ data, the extragalactic sky will already be mapped across a broad range of wavelengths. Deep imaging and spectroscopy from current and upcoming facilities such as Euclid, LSST, DESI, 4MOST, ALMA, JWST, and the Roman Space Telescope will provide detailed measurements of stellar content (masses, structure), star formation across multiple timescales (UV–optical–IR–radio continuum), molecular gas, dust, and hot gas across diverse galaxy populations and environments. SKA \hi\ surveys will therefore be designed to exploit this wealth of legacy and near-future data, linking atomic gas to galaxy growth, baryon cycling, and environmental processes with an unprecedented context. Field selection and survey strategy can now be optimized around these expected datasets, with reliable redshifts remaining essential to extend \hi\ studies below the survey detection threshold and for stacking and environmental analyzes. Survey fields should therefore prioritize overlap with existing or planned wide-area spectroscopic programs, such as DESI and 4MOST (including surveys such as 4HS and WAVES), with additional targeted follow-up where necessary.

For the ultra-deep level, the primary science goals include tracing the evolution of the \hi\ mass function, measuring gas fractions at higher redshift, and linking atomic gas to star formation and galaxy structure in regimes where individual detections are rare. This strongly favors targeting well-established extragalactic deep fields with extensive multi-wavelength coverage and deep spectroscopy. Anchoring the survey to legacy fields with rich ancillary datasets, including deep near- and mid-IR imaging from JWST and wide-field high-resolution imaging from the Roman Space Telescope, maximizes the scientific return in regimes where direct \hi\ detections are sparse and detailed physical interpretation relies heavily on ancillary data.

The medium-deep and wide tiers are optimized for population studies, including gas scaling relations, environmental trends, the connection between atomic gas and star formation, and the identification of rare systems. These surveys will provide large numbers of \hi\ detections, enabling statistical studies of galaxy populations and, for sufficiently nearby or massive systems, resolved characterization of individual galaxies.
To support this science, homogeneous multi-band photometry over large areas (e.g. Euclid, LSST, and Roman) together with wide-area spectroscopy (e.g. 4MOST and DESI) are required to derive robust stellar masses, star formation rates, global galaxy properties, and accurate redshifts for stacking analyzes and environmental measurements over large volumes. More detailed constraints for individual galaxies (such as molecular gas or hot-gas diagnostics) will mainly come from targeted follow-up observations motivated by SKA-Mid \hi\ discoveries.

\subsubsection{Practical considerations}

The possible surveys presented here will face various challenges that careful advance planning and technical developments can help mitigate.  

First, almost every \hi\ survey suffers from RFI, either from solar interference or man-made RFI. Although there is software that can mitigate solar interference \citep[e.g.][]{Samboco_24}, which can help all proposed surveys, it is obviously easier to observe at night as most \hi\ pathfinder surveys have done.  This limits the available time for surveys and makes telescope scheduling trickier.  Unfortunately, the problem is much more challenging regarding man-made RFI.  From Figure~\ref{fig:rfi}, it is clear that large amounts of redshift space are being discarded due to man-made RFI. Although CHILES \citep{Luber_2025} has shown that science can be done at these frequencies, it requires far more careful processing and excision of bad data.  Without improved RFI mitigation strategies, large blocks of cosmic time will be lost to future observations with the SKA.  

The other major challenge will arise for deeper surveys with multiple tracks on a single field (in particular the deep and ultra-deep surveys).  For these surveys, previous experience has shown that the best images arise from joint deconvolution of all calibrated visibilities \citep[e.g.][]{Blyth_16, Luber_2025}.  This leads to working with very large data volumes that require significant computational resources from the SKA Regional Centers or the SKA Regional Center Network (SRC) to image.  This load can be reduced by saving the averaged visibilities of all sessions in a common uv-grid, \citep[as done by CHILES,][]{Luber_2025}.  This further allows for the mitigation of low-level RFI when gridding the data.  The demands of these deep spectral line data at higher dynamic ranges will provide a strenuous test of the SKA data processing and require a significant effort during commissioning.

\section{Summary}

The \hi\ science case presented across the ten chapters of this volume spans the full baryon cycle, from gas accretion and star formation to feedback and quenching, with a compelling continuity across scales and cosmic epochs. It begins with the detailed physics of cloud formation and the atomic-to-molecular transition in the Milky Way and nearby galaxies, extends through the evolution of gas content and scaling relations across galaxy populations over the past several billion years, and reaches to $z\sim3$ and beyond with \hi\ and OH absorption lines offering a dust-unbiased view of cold gas in galaxies. High spectral resolution is a common requirement across emission, absorption, and ISM science, while good surface brightness sensitivity is essential for detecting diffuse low column density gas in galaxy outskirts, the circumgalactic medium, and the Galactic halo. 

Addressing this range of science requires survey programs that span different tiers of depth, area, and spectral resolution, as outlined in different chapters. This overview chapter builds on the tiered survey strategy first proposed in SO15, now updated for AA$^*$ and AA4 configurations to reflect both the updated sensitivity and a more realistic accounting of the RFI environment. The ambition outlined here must be matched by equally serious investment in RFI mitigation strategies and processing infrastructure at the SKA Regional Centers. 

A decade of MeerKAT, ASKAP, uGMRT, FAST and VLA results have demonstrated the feasibility of tracing $\Omega_{\hi}$ to $z\sim1$ but have also exposed tensions between independent measurements that only a large and homogeneous SKA survey can resolve decisively. Direct detections are beginning to open the window beyond $z = 0.5$; the SKA will push this boundary substantially further. Across all these programs, the full scientific return will depend critically on synergy with facilities such as Euclid, LSST, DESI, 4MOST, JWST, and ELTs, where the atomic gas can be placed in the context of ambient physical conditions, stellar mass assembly, star formation history, and large-scale structure.

\bibliographystyle{abbrvnat-maxbibnames4}
\bibliography{chapter} 

\end{document}